\documentclass[twocolumn,prb,reprint,showpacs,preprintnumbers,amsmath,amssymb,superscriptaddress]{revtex4-1}
\usepackage{graphicx}
\usepackage{dcolumn}
\usepackage{bm}
\usepackage{graphicx}
\usepackage{subfigure}
\usepackage{physics}
\usepackage{epstopdf}
\usepackage{braket}
\usepackage{enumerate}
\usepackage{xcolor}

\begin{document}

\title{Eliminating reservoir density-of-states fingerprints in Coulomb blockade spectroscopy}
\author{Arnab Manna}
\affiliation{Department of Electrical Engineering, Indian Institute of Technology Bombay, Mumbai, India\\}
\affiliation{Department of Physics, Indian Institute of Technology Bombay, Mumbai, India\\}
\author{Bhaskaran Muralidharan}
\email{bm@ee.iitb.ac.in}
\affiliation{Department of Electrical Engineering, Indian Institute of Technology Bombay, Mumbai, India\\}
\author{Suddhasatta Mahapatra}
\email{suddho@phy.iitb.ac.in}
\affiliation{Department of Physics, Indian Institute of Technology Bombay, Mumbai, India\\}
\date{\today}

\medskip
\widetext
\begin{abstract}
The differential conductance map of a single electron transistor
(SET) provides information about a variety of parameters related to (quantum) dots, relevant for semiconductor-based quantum computing schemes. However, in ultra-scaled device architectures, identification of excited-state resonances of the quantum dot in the conductance map is often complicated by the appearance of features due to non-uniform density-of-states (DOS) of the source and drain reservoirs of the SET. Here, we demonstrate theoretically that the pump-probe spectroscopic technique, originally introduced by Fujisawa et al.\cite{FujisawaAllowedforbiddentransitions2002,FujisawaTransientcurrentspectroscopy2001a,FujisawaTimedependentsingleelectrontransport2006}, allows the fingerprint of the reservoir-DOS to be completely suppressed, while preserving the visibility of the excited state resonances. We also propose a specific approach for performing DC Coulomb blockade spectroscopy, which can effectively eliminate the DOS-related features. The advantages and limitations of the two approaches are investigated in detail. The results demonstrated here may provide an important optimization capability for emerging proposals of computer-automated control of large arrays of coupled but independently controlled charge and spin qubits.
\end{abstract}
\maketitle
The study of electron transport through a (quantum) dot in the Coulomb Blockade regime, or
Coulomb-blockade spectroscopy, serves as an important first-step in semiconductor-based quantum
computing (QC) implementations. At the single electron limit, quantum dots (QD) represent the
physical qubits in both spin and charge-based QC schemes. On the other hand, a (quantum) dot with
few (hundred) electrons, weakly tunnel-coupled to large electron reservoirs, and capacitively coupled
to a control gate, constitute a single electron transistor (SET). SETs act as extremely sensitive
electrometers, which enable projective measurement of the qubit-state, for both charge and spin qubits\cite{KouwenhovenFewelectronquantumdots2001,HansonSpinsfewelectronquantum2007}. From DC Coulomb-blockade spectroscopy (CBS), information about the
orbital and spin states of the QDs, the electron g-factor, the electron temperature, the strength of the
QD’s tunnel coupling to the source (S) and drain (D) reservoirs, and its capacitive coupling to the
control gates can be determined\cite{nazarovbook}. However, the excited-state resonances of the QD are often
marred in the CBS data by fingerprints of features “extrinsic” to the QD\cite{EscottResonanttunnellingfeatures2010}, particularly by those
due to non-uniform density-of- states (DOS) or local density-of- states fluctuations (LDOSF) in the
source and drain reservoirs.\\
\indent Appearance of such DOS/LDOSF-related features have been reported for a variety of systems, such as
gated single/multi-donor devices\cite{Fuechslesingleatomtransistor2012a,TanTransportSpectroscopySingle2010,LansbergenGateinducedquantumconfinementtransition2008,PierreSingledonorionizationenergies2009,FuechsleSpectroscopyfewelectronsinglecrystal2010a}, MOS-based \cite{LimObservationsingleelectronregime2009} and nanowire-based \cite{ZwanenburgSpinStatesFirst2009} SETs in silicon, and nominally two-dimensional (2D) resonant tunnelling structures fabricated in GaAs/AlGaAs heterosystems \cite{SchmidtSpectroscopylocaldensity1996,SavchenkoResonantTunnellingSpectroscopy2000,FalkoImagelocaldensity1997,HolderEnhancedFluctuationsTunneling2000,SchmidtEnergyDependenceQuasiparticle2001,KonemannCorrelationfunctionspectroscopyinelastic2001,SleightElectronspectroscopicstudyvertical1996}. While these features reveal several interesting information
about the disorder and carrier localization in the leads, they complicate the analysis of the features
related to the QD itself. The challenges involved in probing the QD states in the presence of extrinsic
resonances are discussed in Refs. \onlinecite{EscottResonanttunnellingfeatures2010,Fuechslesingleatomtransistor2012a,TanTransportSpectroscopySingle2010,LansbergenGateinducedquantumconfinementtransition2008,mthesis,MottonenProbecontrolreservoir2010}. Additionally, Ref. \onlinecite{MottonenProbecontrolreservoir2010} suggests one method to
distinguish between the features due to the discrete states of the QD, and those appearing due to the
non-uniform DOS of the quasi-1D S/D leads. However, the approach requires additional gates in the
SET design, to independently control the energy states of the leads.\\
\indent In this Letter, using Coulomb blockade rate equations, we demonstrate that the DOS non-uniformity (or LDOSF) fingerprints may be completely eliminated via the pump-probe technique, originally introduced by Fujisawa et al.\cite{FujisawaAllowedforbiddentransitions2002,FujisawaTransientcurrentspectroscopy2001a,FujisawaTimedependentsingleelectrontransport2006}. Here, the SET current is measured by applying square wave voltage pulses to the top gate (together with a DC component) in the presence of a relatively small S/D bias. We show that the excited-state (ES) resonances are preserved in the data\cite{FujisawaTransientcurrentspectroscopy2001a,FujisawaTimedependentsingleelectrontransport2006}, provided voltage pulses of sufficiently high frequencies can be applied to the plunger gate(s). We then show that DOS-related features may also be suppressed in usual DC Coulomb blockade spectroscopy by judiciously modifying the way in which the S/D bias is applied. Our results suggest that the latter technique is more versatile and possibly easier to implement.\\
\begin{figure}[htb!]	
\includegraphics[width=3in]{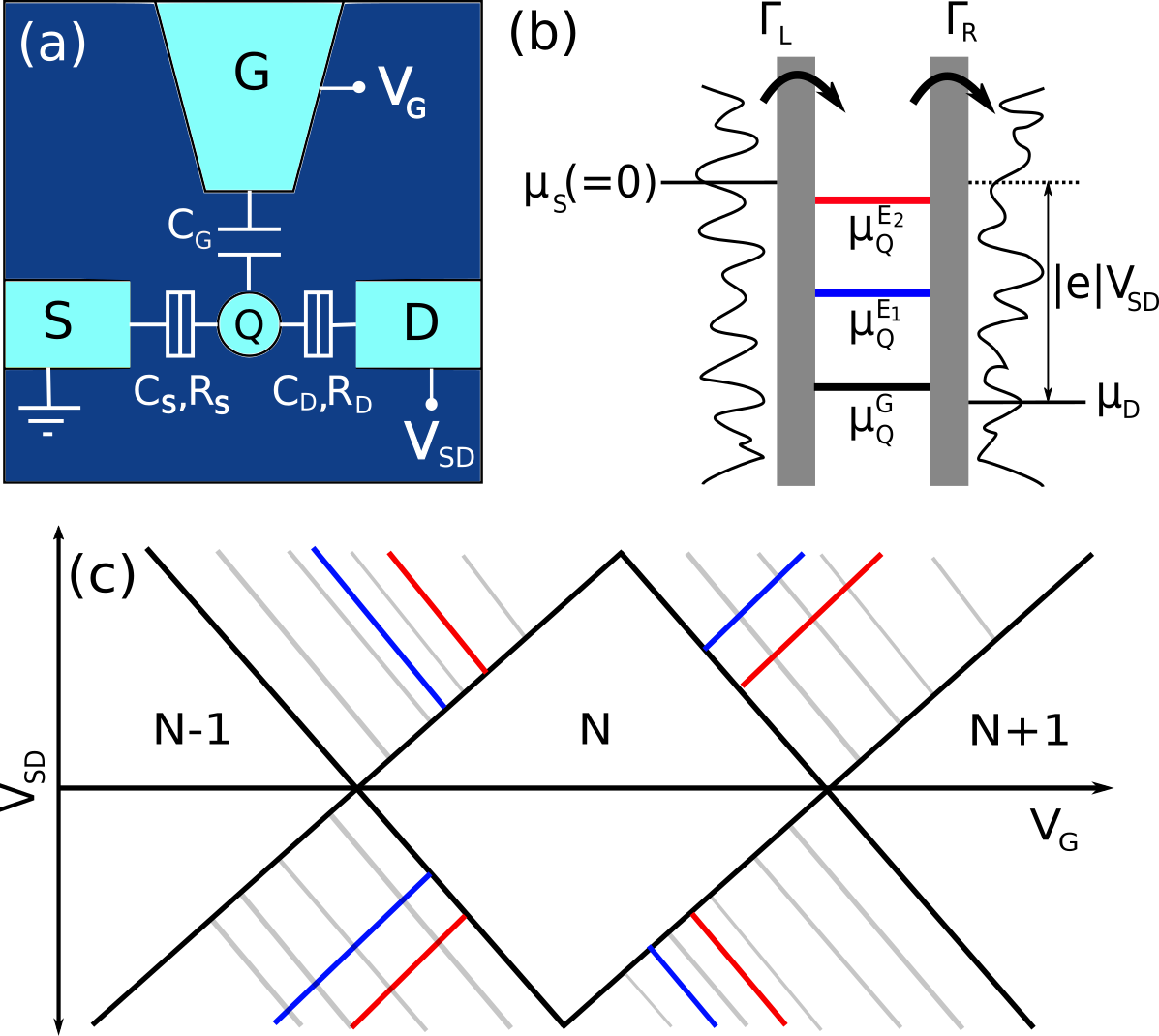}
\caption{(a) Schematic illustration of a SET, depicting the different mutual capacitances and the tunnel-resistances in the circuit. (b) The energy landscape from the source (S) to the drain (D) reservoir, via the quantum dot (Q), showing the non-uniform DOS in the S/D leads and the discrete energy states of Q. (c) Schematic representation of the differential conductance of the SET, revealing the excited state resonances (blue and red lines), along with features due to DOS of the S/D leads (grey lines).}
\label{fig1}
\end{figure}
\begin{figure*}[htb!]	
\includegraphics[width=6in]{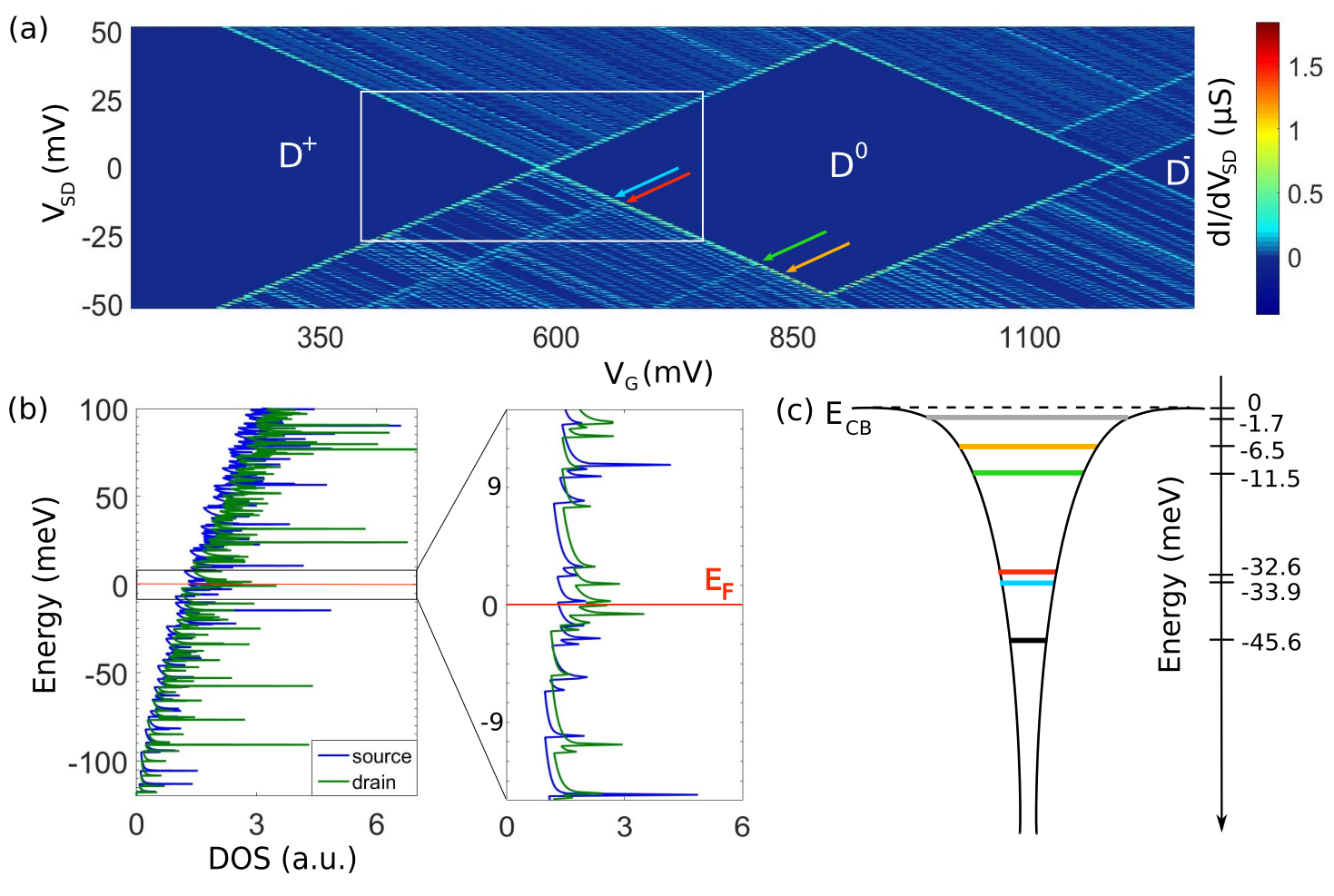}
\caption{(a) Simulated differential conductance map of the single-P-donor SET. (b) Plot of the density of states in the source and drain leads. (c) Schematic illustration of the P-donor potential, depicting the single-electron ($D^0$) and two-electron ($D^-$) ground states, along with the excited $D^0$- states.}
\label{fig2}
\end{figure*}
\indent To elucidate the complications associated with the appearance of DOS/LDOSF features in DC CBS data, we consider the SET circuit shown schematically in Figure~\ref{fig1}(a). The QD (labeled Q) is tunnel-coupled to the S/D leads (characterized by the capacitances $C_S$  and $C_D$, and tunnel-resistances $R_S$ and $R_D$) and only capacitively coupled to the plunger gate (G), (characterized by the capacitance $C_G$). Mutual capacitances between the S/D leads and G have been ignored for the sake of simplicity. Figure~\ref{fig1}(b) depicts the energy landscape of the SET under an applied source-drain bias ($V_{SD}$). The voltage ($V_G$) on the plunger gate controls the electrochemical potential of the dot ($\mu_Q$), with respect to that of the reservoirs ($\mu_S$  and $\mu_D$). Together with the discrete energy states of the QD, the continuum of energy states in the S/D leads, characterized by a non-uniform energy-dependent DOS (or LDOSF), is also sketched in Fig.~\ref{fig1}(b). In the corresponding schematic 2D-map of the differential conductance of the SET (Figure~\ref{fig1}(c)), which contains diamond-shaped Coulomb-blockaded regions with well-defined electron count on the dot (N), the fingerprint of the DOS non-uniformity/LDOSF is depicted as (grey) lines parallel to one of the diamond edges. These fluctuations of the SET conductance appear in the CBS data since the discrete energy state of the QD acts as an efficient spectrometer for probing the DOS/LDOSF in the leads, when it is plunged through the $V_{SD}$ bias window, by varying $V_G$ \cite{SchmidtSpectroscopylocaldensity1996}. The resolution of the spectrometer is only limited by the level broadening of the QD energy states, which depends on the strength of the tunnel-coupling between the QD and the S/D leads and/or the electron temperature. Thus, at milli-Kelvin temperatures and for weak tunnel-coupling, typical of SETs used as sensitive electrometers, the DOS/LDOSF features can be very well-resolved, which makes identification and analysis of excited-state resonances of the QD (depicted by blue and red lines in Fig.~\ref{fig1}(c)) challenging.\\
\indent To model the conductance map of DC Coulomb blockade spectroscopy, as well as to demonstrate the results of the pump-probe spectroscopy, we consider a realistic SET lay-out, similar to that reported in Ref. \onlinecite{Fuechslesingleatomtransistor2012a}. Here, the entire SET is fabricated on a single atomic plane within a Si crystal, by selected-area phosphorus doping, using a scanning-tunnelling-microscopy (STM)-based lithography technique. The QD of the SET is defined by a single P donor, incorporated between the S and D leads, similar to the design schematically shown in Fig.~\ref{fig1}(a).  For the mutual capacitances in the SET layout, we assumed experimentally obtained values \cite{mthesis} for the same device. \\
\indent For the steady state transport through the SET in the Coulomb blockade regime, the addition and removal of electrons is described by a rate equation\cite{MuralidharanProbingelectronicexcitations2006,MuralidharanGenericmodelcurrent2007} for the non-equilibrium probability $P_i^N$ of each N-electron many-body state $|N,i>$ with total energy $E_i^N$. The master equation involves transition rates $R_{(N,i)\rightarrow(N\pm 1,j)}$  between states differing by a single electron, leading to a set of independent equations, defined by the size of the Fock space:
\begin{align}
\frac{dP_i^N}{dt}= -\sum_j[R_{(N,i)\rightarrow(N\pm 1,j)} P_i^N - R_{(N\pm 1,j)\rightarrow(N,i)} P_j^{N\pm 1} ] + R_{r}   
\label{eq1}
\end{align}
and the normalization equation,  $\sum_{i,N}P_i^N = 1$. $R_{r}$ is the relaxation term given by:
\begin{align}
R_{r}= 
\begin{cases}
    \frac{\sum_{j\neq i} R_{(N,j)\rightarrow(N,i)}}{\tau_r} P_i^N,& \text{if } i = GS\\
    - \frac{1}{\tau_r} P_i^N,              & \text{otherwise}
\end{cases}
\end{align}
where $\tau_r$ is the relaxation time constant. The transition rates are given by:
\begin{align}
R_{(N,i)\rightarrow(N-1,j)}=\sum_{\alpha=S,D}\Gamma_\alpha [1-f(\mu_{ij}^{Nr} - \mu_\alpha )], \\R_{(N-1,j)\rightarrow(N,i)}=\sum_{\alpha=S,D}\Gamma_\alpha f(\mu_{ij}^{Na} - \mu_\alpha )
\label{eq2}
\end{align}
where $\mu_\alpha$ are the electrochemical potentials of the S/D contacts, and $f$ is the corresponding Fermi function, with single particle removal and addition
transport channels $\mu_{ij}^{Nr}=\mu_i^N - \mu_j^{N-1}$ and $\mu_{ij}^{Na} = \mu_j^{N+1}-\mu_i^N$, respectively. Here, $\Gamma_\alpha$ denote the tunnel rates, which contain the expression of DOS of the respective leads. In steady state, $(\dfrac{dP_i^N}{dt}=0)$, the current through the SET ($I_{SD}$ ) can be equated to the current across one of the barriers as
\begin{align}
I_{SD} &= I_S = -I_D \nonumber \\
&=\pm \frac{e}{\hbar} \sum_{ij}[R_{(N,i)\rightarrow(N\pm 1,j)}^S P_i^N - R_{(N\pm 1,j)\rightarrow(N,i)}^S P_j^{N\pm 1}]  
\label{eq3}
\end{align}
\indent Figure~\ref{fig2}(a) shows the simulated differential conductance map for the device. Here, the tunnel coupling of the QD to the S/D leads is assumed to be symmetric (with tunnel rates $\Gamma_S$= $\Gamma_D$=100 GHz\cite{TettamanziProbingQuantumStates2017}), while the energy relaxation rate from the ES to the ground state GS is taken to be W=$\tau_r^{-1}$=100 MHz\cite{TahanRelaxationexcitedspin2014a,VolkProbingrelaxationtimes2013}. The map reflects the positively-charged ($D^+$), neutral ($D^0$), and negatively-charged ($D^-$) charge states of the P-donor\cite{Fuechslesingleatomtransistor2012a}.  The potential profile for an isolated P donor in bulk Si is shown in Figure~\ref{fig2}(c). The $D^0$ and $D^-$ ground states (GS) are situated below the Si conduction band edge ($E_{CB}$ ) by their respective binding energies (45.6 meV and 1.7 meV). Also shown are the $D^0$ ES, together with their energy separation from the ground state. In the $D^0$ diamond of Fig.~\ref{fig2}(a), the spectra of the excited states (marked by the arrows) appear at $V_{SD}$  values which correspond to their energies, as shown in Fig.~\ref{fig2}(c). However, superposed on this spectrum are additional resonances, which are attributable to the energy-dependence of the DOS of the S and D leads. \\
\indent To model the DOS in the S/D leads (Fig.~\ref{fig2}(b)), we have considered an ideal DOS of a 70 nm wide wire\cite{TanKuanthesis}. Although this simplification possibly underestimates the randomness of the DOS energy-dependence, by neglecting the non-regular geometry of the actual S/D leads and any LDOSF, the observations related to the study of their appearance in the CBS data are expected to remain the same. It is also important to note that, though we demonstrate the results of our modelling for the device design of Ref. \onlinecite{Fuechslesingleatomtransistor2012a}, the conclusions of this work may be generalized for any SET lay-out. Figure~\ref{fig2}(a) clearly demonstrates that the DOS related features make the identification of the resonances due to the donor ES complicated, and in many cases impossible \cite{PierreSingledonorionizationenergies2009,FuechsleSpectroscopyfewelectronsinglecrystal2010a,ZwanenburgSpinStatesFirst2009}.  \\     
\begin{figure}[htb!]	
\includegraphics[width=3.4in]{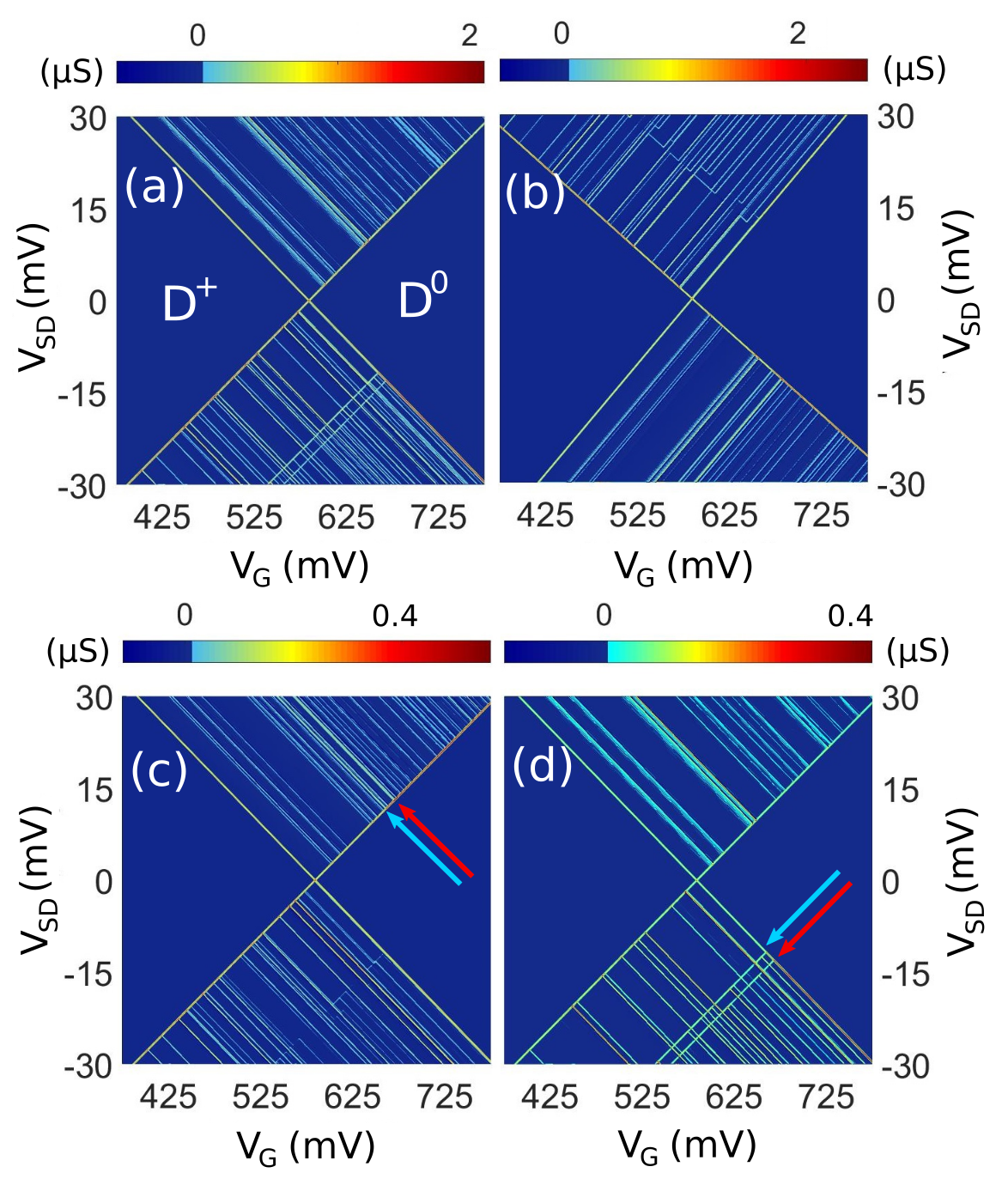}
\caption{Differential conductance maps simulated around the $D^0\leftrightarrow D^-$ transition (a) for $\Gamma_S= \Gamma_D=100$ GHz, $V_{SD}$ applied to the D terminal, (b) for $\Gamma_S= \Gamma_D=100$ GHz, $V_{SD}$ applied to the S terminal, (c) for $\Gamma_S=100$ GHz, $\Gamma_D=10$ GHz, $V_{SD}$ applied to the D terminal, and (d) for $\Gamma_S=10$ GHz, $\Gamma_D=100$ GHz, $V_{SD}$ applied to the D terminal.}
\label{fig3}
\end{figure} 
\indent Characteristic features of the DOS spectrum in DC CBS data are further explained in Figure~\ref{fig3}. Figure~\ref{fig3}(a) shows a close up image of the region demarcated by the rectangle in Fig.~\ref{fig2}(a). Here, $V_{SD}$ is applied to D, while S is grounded (See representation of Fig.~\ref{fig1}(b)).  The DOS resonances in this case run parallel to the Coulomb-diamond-edge with negative slope $\dfrac{dV_{SD}}{dV_G}<0$. However, when $V_{SD}$ is applied to S (with D grounded), the DOS resonances run parallel to the Coulomb-diamond-edge with positive slope $\dfrac{dV_{SD}}{dV_G}>0$, as shown in Figure~\ref{fig3}(b). We note that this finding is in contrast to claims made in several previous reports\cite{EscottResonanttunnellingfeatures2010,mthesis,MottonenProbecontrolreservoir2010}, where the DOS resonances have been schematically shown and argued to run parallel to both diamond edges, for $V_{SD}$ applied to either of the left/right terminals. According to our survey, only Ref. \onlinecite{LansbergenGateinducedquantumconfinementtransition2008} makes a correct depiction of the DOS resonances. \\
\indent In case of significantly asymmetric S/D tunnel coupling ($\Gamma_S \neq \Gamma_D$), $I_{SD}$ is determined by the slower tunnel-rate \cite{FujisawaTimedependentsingleelectrontransport2006}. Therefore, features due to only the DOS of the weaker-coupled lead appears in the conductance map. Figure~\ref{fig3}(c) and 3(d) depict the cases where $\Gamma_D=0.1\Gamma_S$  and $\Gamma_S=0.1\Gamma_D$, respectively. In comparison to Fig.~\ref{fig3}(a) ($\Gamma_S=\Gamma_D=100$ GHz), fewer DOS- resonances are observed in both cases. However, it is also evident that the QD ES-transitions are pronounced only on the $V_{SD}>0$ ($V_{SD}<0$) half of the conductance map, in Fig.~\ref{fig3}(c) (Fig.~\ref{fig3}(d)). For the other half of the corresponding maps, electron extraction (injection) takes place on a slower (faster) timescale, thus making relaxation to GS more likely. This can complicate the analysis, albeit it can be mitigated to some extent by judiciously choosing the lead, to which $V_{SD}$ is applied.\\
\indent In contrast to the DC Coulomb blockade spectroscopy, the pump-probe spectroscopy (PPS) technique allows complete elimination of the DOS-resonances\cite{TettamanziProbingQuantumStates2017}. In PPS, a rectangular voltage pulse $V(t)$, is applied to the gate (Figure~\ref{fig4}(b)), together with a DC voltage, and $V_{SD}$ is maintained small.  Here, the time-integrated non-equilibrium transient current is mapped out as a function of the DC gate voltage ($V_G$) and the amplitude of the rectangular pulse ($|V_p|$), as originally introduced by Fujisawa et al. \cite{FujisawaTransientcurrentspectroscopy2001a}. A schematic of the current ($I_{SD}$)-map is shown in Figure~\ref{fig4}(a).  The DC Coulomb peak (i.e. for $V(t)=0$) corresponding to the $GS(N-1)\leftrightarrow GS(N)$ transition appears at a particular value of $V_G (=V_G^0)$ . On the other hand, for $|V_p|>0$, the same transition occurs at two different values of $V_G (=V_G^L$ and $V_G^R)$, which are equidistant from the $V_G=V_G^0$ line (See Fig.~\ref{fig4}(a)). To explain this current map, we first consider the situation at $V_G=V_G^L$. In the low-level of the voltage pulse $(V(t)=-V_p)$, both GS and ES electrochemical potentials, $(\mu_Q^G (N)$  and $\mu_Q^E (N))$, are well above the $V_{SD}$ window, so that the SET is in the Coulomb blockade regime, with  N-1 electrons on the QD. When $V(t)$ is ‘instantaneously’ increased to $+V_p$,  both $\mu_Q^G (N)$ and $\mu_Q^E (N)$ are plunged down. The SET current, $I_{SD}$, is switched on once at  $V_p=+V_p^G$, when $\mu_Q^G (N)$ is plunged down into the $V_{SD}$ window (with $\mu_Q^E (N)$ situated above the window), and again at $V_p=+V_p^E$, when $μ_Q^E (N)$ is plunged down into the $V_{SD}$ window (with $\mu_Q^G (N)$ situated below the window). In case of the later, electron transport takes place as long as the $\mu_Q^G (N)$ remains unoccupied. On the other hand, at $V_G=V_G^R$, $\mu_Q^G (N)$  and $\mu_Q^E (N)$ are situated well below the $V_{SD}$ window for $V(t)=+ V_p$ (Coulomb blockade regime, with  N electrons)  and plunged up when $V(t)$ is instantaneously decreased to $- V_p$. Here, the SET current switches on only for $V_p=- V_p^{(G)}$, when $\mu_Q^G (N)$ is plunged up into the $V_{SD}$ window. For a larger magnitude of $|V_p|$, the transitions take place at values of $V_G$ farther away from $V_G^0$, explaining the characteristic V-shape for the GS transition and the slope of the ES resonance.\\
\indent Although the current map of the pump-probe technique resembles that of DC CBS, there are several important differences. The energy of the ES (above the GS) cannot be directly read off from the vertical axis of the pump-probe current map (unlike the case of a DC CBS current/differential-conductance map). This is because the voltage $V(t)$ is applied to the gate terminal, which yields the energy of the ES only when the correct lever arm $(\alpha_G = \frac{C_G}{C_\Sigma})$ is accounted for. Secondly, the slope of the edges of the V-shape are independent of the mutual capacitances of the SET circuit, unlike that of the diamond edges in DC CBS (ideally, $\frac{dV_G^P}{dV_G^{DC}})=1$). Finally, as demonstrated next, the DOS features do not appear in the pump-probe current map. \\
\begin{figure*}[htb!]	
\includegraphics[width=6.6in]{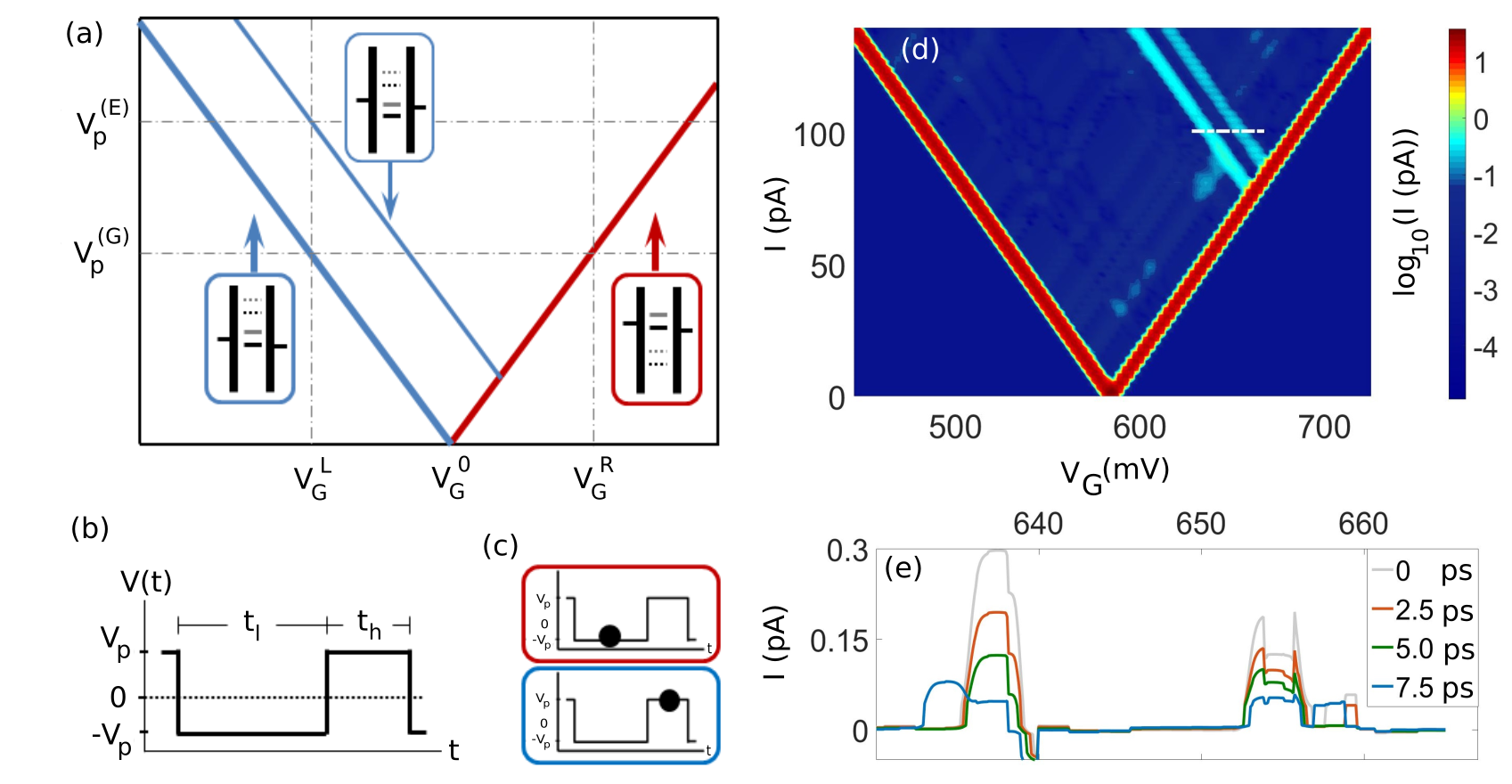}
\caption{Schematic illustration of (a) current map recorded in pump-probe spectroscopy (PPS), (b) the voltage pulse function, and (c) the pulse instances corresponding to the 'blue' and 'red' lines in (a). (d) Simulated current map of PPS. (e) Plot showing the excited-state current for different rise times along the white dashed line in (d).}
\label{fig4}
\end{figure*} 
\indent For the single-P-donor SET considered earlier, the time-integrated current of the pump-probe technique:
\begin{align}
I = \frac{1}{T}\sum_{t=0}^T \frac{I_S(t)-I_D(t)}{2}
\end{align}
is simulated by obtaining the non-steady state solution to Equation\eqref{eq1}, using a finite-difference backward Euler scheme. The simulated current map (with $\Gamma_S= \Gamma_D=100 GHz$, and $W=\tau_r^{-1}=100$ MHz), for a range of $V_G$ around the $D^+\leftrightarrow D^0$ transition, is shown in Figure~\ref{fig4}(d). The first two ES resonances of the $D^0$-state can be seen in the map, without the interference of features due to the DOS of the S/D leads. It is easy to comprehend why the DOS features vanish from the pump-probe data. Since the $\mu_S$  and $\mu_D$ are kept fixed, the ‘spectrometer’ ($\mu_Q$) cannot scan the DOS of either of the leads. Even temporal-fluctuations of DOS will only modulate the current along the V-shape, but will not result in additional resonances in the current map. \\
\indent It is also important to note the conditions under which the ES-resonances remain visible in the PPS technique. Electron tunnel rates across S/D barriers, can range from 1 MHz to 100 GHz in typical SETs \cite{FujisawaTimedependentsingleelectrontransport2006}. In contrast, momentum relaxation rates are typically $W\geq 10$ MHz \cite{TettamanziProbingQuantumStates2017}. For the PPS technique to reveal the orbital ES-resonances, the pulse-high time ($t_h$), needs to satisfy the condition $W<t_h^{-1}<\Gamma_S,\Gamma_D$. While $t_h^{-1}<\Gamma_S,\Gamma_D$ ensures that within each cycle of the voltage pulse, $\mu_Q^E (N)$ resides for sufficient time within the bias window for  measurable electron tunnelling processes to occur, $t_h^{-1}>W$ warrants that orbital relaxation does not populate the GS, and thereby, block the SET current. Thus voltage pulses of at least a few MHz are typically required to record ES-resonances in PPS. On the other hand, the spin relaxation rate in silicon is $\sim$ Hz. Hence, identification of the spin ES of a single electron confined to a QD can be amply performed by the PPS technique, even for weak tunnel coupling to the S/D leads. In real devices, often the S/D tunnel-coupling to the QD is highly asymmetric.  For such cases, electron injection (extraction) across the slower (faster) barriers is an additional requirement for good visibility of the ES-resonances.\\ 
\indent Another requirement for the visibility of the ES-resonances is that the rise time of the pulse is much shorter than $\Gamma_{S,D}^{-1}$. Figure~\ref{fig4}(e) plots $I_{SD}$  for a section of the ES-resonances in Fig.~\ref{fig4}(a) (shown by the dotted white line), for different pulse rise-times (corresponding to $\Gamma_S= \Gamma_D=100$ GHz and $\tau_h^{-1}= 20$ GHz). It is observed that the current corresponding to the ES-resonances falls sharply with increasing rise-time. For a higher rise-time, electron-tunnelling to the GS (situated below the $V_{SD}$ window) becomes more likely, which blocks the current through the ES (situated within the $V_{SD}$ window), due to Coulomb blockade. \\
\indent While ES-resonances can be resolved in the PPS data, without being obfuscated by the DOS fingerprint, their visibility is contingent upon an intricate balance between the different rates. A favourable ratio of $W$ and $\Gamma_{S,D}$ may not be always accessible, either due to constraints imposed by the device architecture or the intrinsic timescales of the (energy) relaxation processes. In this section, we demonstrate that the DOS features may also be eliminated from the differential conductance data of the usual DC CBS, by sweeping the bias voltage on both S and D terminals, in a ratio determined only by the mutual capacitances $C_S$ and $C_D$. The electrochemical potential ($\mu_{ij}^N$) of the QD is related to the voltages applied to the different terminals as \cite{HansonSpinsfewelectronquantum2007}:
\begin{align}
\mu_{ij}^N = (N-1/2)E_C - \frac{E_C}{|e|}(C_S V_S + C_D V_D + C_G V_G ) + \Delta E    
\label{eq5}
\end{align}
\begin{figure}[htb!]	
\includegraphics[width=3.4in]{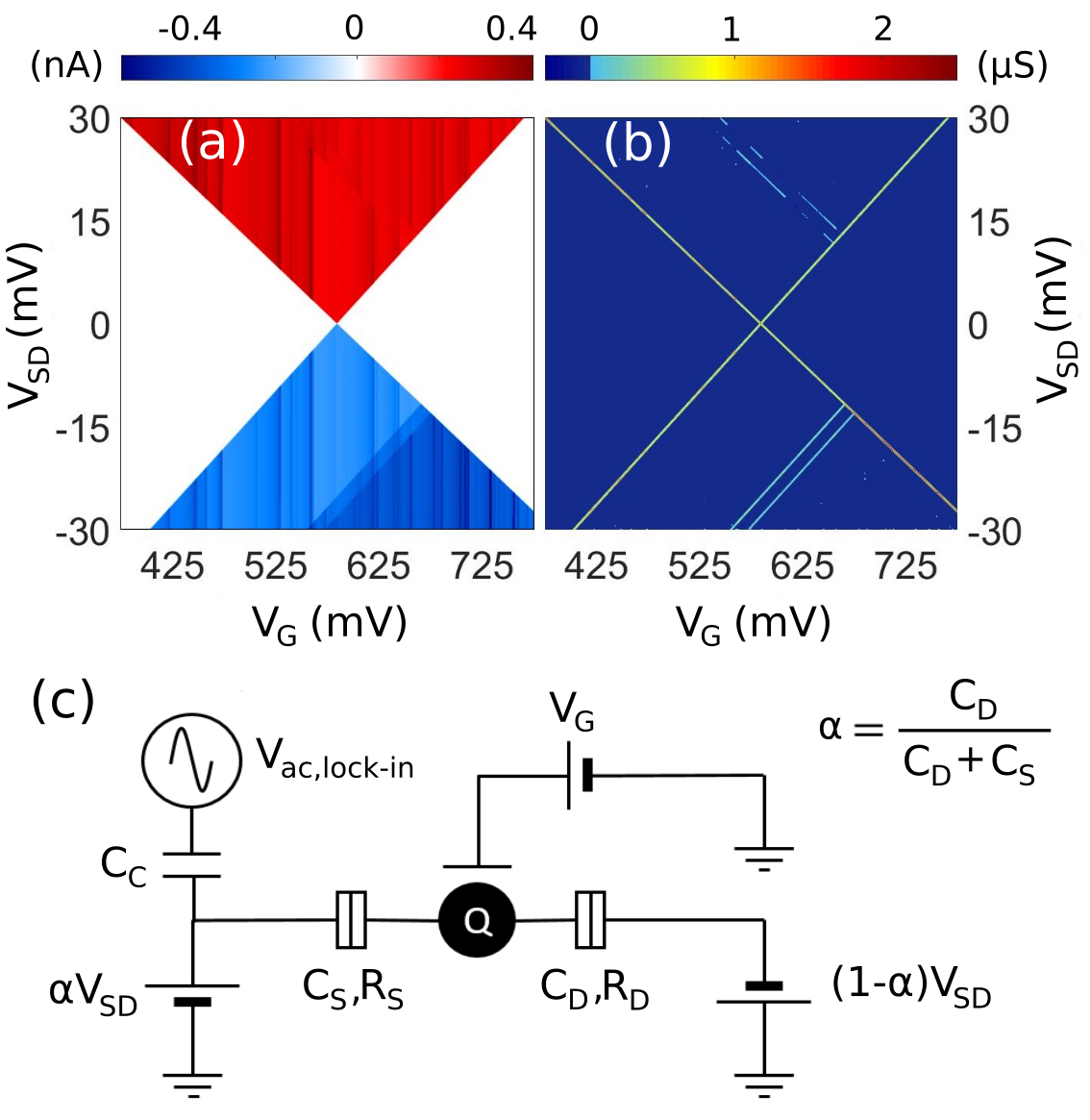}
\caption{The simulated (a) current and the (b) differential conductance maps for the DC biasing scheme shown in (c). }
\label{fig5}
\end{figure} 
Here, $E_C=e^2/(C_S+C_D+C_G )$ is the charging energy of the QD, $\Delta E=E_j^N-E_i^{N-1}$ is the difference between the eigenenergies of the $|N,j>$ and the $|N-1,i>$ states, and $V_S,V_D,V_G$ are the voltages applied to the S, D, and G terminals, respectively. If the S/D-bias is applied such that $V_S=\alpha V_{SD}$ and $V_D=-(1-\alpha) V_{SD}$ (See Figure~\ref{fig5}(c)), where $\alpha=C_D/(C_S+C_D)$, $\mu_{ij}^N$ will depend only on $V_G$. In such a scenario, the DOS-resonances in the current map will appear as lines parallel to the $V_{SD}$-axis, as shown in Figure~\ref{fig5}(a). In the corresponding differential conductance map (Figure~\ref{fig5}(b)), the DOS-signature is completely suppressed. We believe that this method is easier to implement and less-constrained than the PPS technique, as the current through the ES is not restricted by the gate voltage pulse-rate or blockaded by the loading of the GS.  \\
\indent In conclusion, we demonstrated in this work that excited-state resonances of quantum dots can be obtained without interference of features due to the non-uniform density-of-states of the source and drain reservoirs in pump-probe spectroscopy and also in usual Coulomb blockade spectroscopy, provided the source-drain bias (in the latter case) is applied is a particular ratio to both terminals. The results bode well for computer-automated initialization\cite{baart_computer-automated_2016} of large array of spin/charge qubits.\\
\textit{Acknowledgements}: AM and SM acknowledge useful discussions with Aritra Lahiri and Sven Rogge, respectively.

\bibliographystyle{apsrev}

\end{document}